\documentclass[3p,times,twocolumn]{elsarticle}

\usepackage{ecrc}

\volume{00}

\firstpage{1}

\journalname{Nuclear Physics B Proceedings Supplement}

\runauth{C. A. Sierra and Y. Rodr\'iguez}

\jid{nuphbp}

\jnltitlelogo{Nuclear Physics B Proceedings Supplement}

\usepackage{amssymb}
\usepackage{verbatim} 
\usepackage{graphicx}
\usepackage{dcolumn}
\usepackage{bm}

\usepackage[figuresright]{rotating}

\begin{document}




\dochead{PI/UAN-2015-587FT}

\title{Moment transport equations and their application to the perturbed universe}

\author[label1]{Carlos A. Sierra}
\ead{carlos.sierra@correo.uis.edu.co}
\author[label1,label2]{Yeinzon Rodr\'{\i}guez}
\ead{yeinzon.rodriguez@uan.edu.co}
\address[label1]{Escuela de F\'isica, Universidad Industrial de Santander, Ciudad Universitaria, Bucaramanga 680002, Colombia}
\address[label2]{Centro de Investigaciones en Ciencias B\'asicas y Aplicadas, Universidad Antonio Nari\~no, Cra 3 Este \# 47A -15, Bogot\'a D.C. 110231, Colombia}


\begin{abstract}
There are many inflationary models that allow the formation of the large-scale structure of the observable universe. The non-gaussianity parameter $f_{NL}$ is a useful tool to discriminate among these cosmological models when comparing the theoretical predictions with the satellite survey results like those from WMAP and Planck. The goal of this proceeding contribution is to review the moment transport equations methodology and the subsequent calculation of the $f_{NL}$ parameter.
\end{abstract}

\maketitle

\section{\label{secA}Introduction}

The origin and evolution of the Universe at large scales can be studied by means of two elements: the first one describes a universe completely homogeneous and isotropic, which is the one usually studied with a flat background metric; the second one describes the cosmological perturbations of the physical quantities that can be measured nowadays, like $\rho$ (energy density), $T$ (temperature), $K$ (spatial curvature), $P$ (pressure), etc. The inflationary mechanism generates isotropic expansion in a space region, and convert the quantum fluctuations into classical perturbations once they are outside the horizon; this, in turn, allows the generation of the large-scale structures \cite{Linde,Liddle,Weinberg}. Those perturbations will evolve thanks to the gravitational accretion to today's celestial bodies. The primordial curvature perturbation $\zeta$ is a fundamental quantity that allows us to obtain the observable parameters of a cosmological model. The non gaussianity of this perturbation, parameterized by the quantity $f_{NL}$, has become one the most important discriminators among inflationary models.

The objective of this proceedings contribution is to review an standardization procedure that allows us to calculate the parameter $f_{NL}$, related to a cosmological model, using the moment transport equations \cite{Mulryne:2009kh,Mulryne:2010rp,Mulryne:2013oba} and some adequate gauge transformations. The motivation of this study relies on the dificulty at obtaining the non-gaussianity parameters of cosmological models by means of the $\delta N$ formalism \cite{Sasaki:1995aw,Lyth:2004gb,Lyth:2005fi}, because there does not exist a stardard analytic procedure to obtain the derivatives of the amount of expansion or number of e-folds $N$. To overtake this difficulty, the use of the moment transport equations is proposed so that the difficulty at obtaining the $N$ derivatives is trade by a direct and standardized way of calculating the evolution of the correlators of the field perturbations.

\section{\label{secB}Standard cosmology and the separate universe assumption}

The standard cosmology employs the cosmological principle \cite{Linde,Liddle,Weinberg} that assumes homogeneity and isotropy in the matter and energy distribution for distance scales bigger than $10$ MPc. The absence of perturbations in quantities like the energy density and pressure, called background quantities, cannot explain the origin and evolution of the actual structure of our universe; indeed, the structure is the result of small fluctuations of these quantities. In the search for an inflationary model that could explain the observations, it is necessary to perturb those quantities to recreate the large-scale structure of our observable universe. 

The $\delta N$ formalism \cite{Linde} is a sharp tool to reduce efforts at calculating cosmological parameters associated with perturbations in inflationary scenarios.  To begin, the expansion of the universe and the fact that the universe is almost flat \cite{Hinshaw:2012aka,Ade:2013ydc}, allow us to propose the Friedmann-Robertson-Walker (FRW) metric in cartesian coordinates as:
\begin{equation}
 g_{\mu\nu} =
\left(
  \begin{array}{cccc}
    -1 & 0  \\
    0 & a^{2}\delta_{ij} \\
  \end{array}
\right),
\end{equation}
where the rate of the expansion is measured through the Hubble parameter $H$ defined by means of the scale factor $a(t)$:
\begin{equation}
H(t)=\frac{\dot{a}(t)}{a(t)},\label{secB_1}\\ 
\end{equation}
where the dot means a derivative with respect to the cosmic time.
The equation that describes the homogeneous and isotropic expansion of the universe (assuming Einstein gravity) is the Friedmann equation \cite{Liddle,Weinberg}:
\begin{equation}
H^2(t) = \frac{\rho(t)}{3m_p^2},\label{secB_2}\\
\end{equation}
where $m_p$ is the reduced Planck mass.

Assuming the cosmological principle, it is possible to express the energy-momentum tensor by means of a perfect fluid, and for cosmological scales (bigger than $10$ MPc) the spatial gradients of the fields vanish \cite{Linde}, that is why it is possible to express the energy density and the pressure in terms of the scalar fields present during inflation in the following way:
\begin{eqnarray}
\rho(t) &=& \frac{1}{2}\Sigma_i^N \dot{\varphi_i^2}+V(\varphi_1(t),\varphi_2(t),..,\varphi_N(t)),\label{secB_3} \\
P(t) &=& \frac{1}{2}\Sigma_i^N \dot{\varphi_i^2}-V(\varphi_1(t),\varphi_2(t),..,\varphi_N(t)).\label{secB_4}
\end{eqnarray}

The continuity equation for the energy density comes from the adiabatic principle applied to volumes bigger than (10 MPc)$^3$, i.e., assuming that there is no heat transfer between those spatial regions:
\begin{equation}
\dot\rho(t)+3H(t)\left(\rho(t)+P(t)\right)=0,\label{secB_5}
\end{equation}
so that, employing Eqs. (\ref{secB_3}) and (\ref{secB_4}), the equations of motion for the scalar fields are obtained:
\begin{equation}
\ddot\varphi_i(t)+3H\dot\varphi_i(t)+\frac{\partial{V}}{\partial{\varphi_i(t)}}=0.\label{secB_6}
\end{equation}

Now, it is time to introduce the separate universe assumption. It states that the universe could be analysed at each point as if it were homogeneous and isotropic near the space point of interest \cite{Linde}. This process is equivalent to smooth out the small-scale details, allowing to rewrite the Eqs. (\ref{secB_1}) - (\ref{secB_6}) as functions of the position: 

\begin{eqnarray}
H(\vec{x},t) &=& \frac{\dot{a_1}(\vec{x},t)}{a_2(\vec{x},t)},\label{secB_7} \\ 
H^2(\vec{x},t) &=& \frac{\rho(\vec{x},t)}{3m_p^2},\label{secB_8}
\end{eqnarray}
\begin{eqnarray}
\rho(\vec{x},t) &=& \frac{1}{2}\Sigma_i^N \dot{\varphi_i}^2\left(\vec{x},t\right)+V(\varphi_1(\vec{x},t),..,\varphi_N(\vec{x},t)), \nonumber \\
&& \label{secB_9}\\
P(\vec{x},t) &=& \frac{1}{2}\Sigma_i^N \dot{\varphi_i}^2(\vec{x},t)-V(\varphi_1(\vec{x},t),..,\varphi_N(\vec{x},t)), \nonumber \\
&& \label{secB_10}
\end{eqnarray}
\begin{eqnarray}
\dot\rho(\vec{x},t)+3H(\vec{x},t)\left(\rho(\vec{x},t)+P(\vec{x},t)\right) &=& 0,\label{secB_11} \\
\ddot\varphi_i(\vec{x},t)+3H(\vec{x},t)\dot\varphi_i(\vec{x},t)+\frac{\partial{V}}{\partial{\varphi_i(\vec{x},t)}} &=& 0.\label{secB_12}
\end{eqnarray}
The Hubble parameter from Eq. (\ref{secB_7}) is defined in general by means of the quantities $a_1$ and $a_2$ evaluated in different slicings\footnote{The slicing corresponds to a set of hypersurfaces with constant time that folds the spacetime. The threading, in contrast, corresponds to a set of hypersurfaces with constant spatial coordinates.}\cite{Linde}.

\section{\label{secC}The primordial curvature perturbation $\zeta$ and the $\delta N $ formula}
After the separate universe idea has been assumed, the scale parameter $a(t)$ must be defined at each point $\vec{x}$:
\begin{equation}
a(\vec{x},t)=a(t)e^{\zeta(\vec{x},t)}.\label{secC_1}\\ 
\end{equation}
This definition of the local scale parameter is composed by two parts: the first part is the global scale factor, and the second one encodes the intrinsic curvature of the perturbed spacetime\footnote{The threading must be comoving and the slicing must be the one of uniform energy density.}.  $\zeta(\vec{x},t)$ is called the primordial curvature perturbation.

Using the definition in Eq. (\ref{secC_1}), the continuity equation from Eq. (\ref{secB_11}), and the local definition of the Hubble parameter in Eq. (\ref{secB_7}), we obtain:
\begin{equation}
\dot{\rho}(t)+3\left[H(t) + \zeta(\vec{x},t) \right]\left[\rho(t) + P(\vec{x},t)\right]=0. \label{secC_2}
\end{equation}
It is worth mentioning that the quantity $\zeta(\vec{x},t)$ is conserved only if the adiabatic condition $P=P(\rho)$ is accomplished \cite{Linde}; this is satisfied, for example, when the universe is in the radiation dominated era $(P=\frac{\rho}{3})$ or in the matter dominated era $(P=0)$.

The global amount of expansion $N(t)$ defined as (see Refs. \cite{Linde,Liddle,Weinberg}):
\begin{equation}
N(t_1,t_2) = ln\frac{a(t_2)}{a(t_1)} , \label{secC_3}
\end{equation}
is easily generalized to the local version:
\begin{equation}
N(\vec{x},t_1,t_2) = ln\frac{a(\vec{x},t_2)}{a(\vec{x},t_1)} , \label{secC_4}
\end{equation}
where the slicings at $t_1$ and $t_2$ must be explicitly stated (they need not be necessarily the same). Thus, remembering the definition in Eq. (\ref{secC_1}), we conclude that:
\begin{equation}
\zeta( \vec{x},t_2) = N(\vec{x},t_1,t_2) - N(t_1,t_2), \label{secC_6}
\end{equation}
where the slicing in $t_2$ is the one of uniform energy density and the slicing in $t_1$ is flat (the scale factor does not have a local part). The latter equation does not depend on the initial time $t_1$ \cite{Linde}. A Taylor-series expansion leads to:\\
\begin{eqnarray}
\zeta(\vec{x},t_2) &=& N_i \ \delta\varphi_i(\vec{x},t_1) \nonumber \\
&+& \frac{1}{2}N_{ij} \ \delta\varphi_i(\vec{x},t_1) \delta\varphi_j(\vec{x},t_1) + \cdots  , \label{secC_8}
\end{eqnarray}
where the scalar field perturbations are defined in the flat slicing and $N_i$, $N_{ij}$, etc., are the derivatives of the \emph{local} $N$ with respect to the scalar fields present during inflation and evaluated in the background. Though $\zeta$ does not depend on the initial time $t_1$, the $N$ derivatives depend both on the initial and final times $t_1$ and $t_2$. However, even if $t_1$ and $t_2$ are made equal, that does not imply that $\zeta$ vanishes;  the crucial aspect here is that, as stated before, the $N$ derivatives are first obtained with the \emph{local} version of $N$ and then evaluated in the background, providing this way a result that is different to zero even when the \emph{global} $N$ vanishes\footnote{All of this can be restated by saying that the curvature perturbation can be calculated by means of an adequate gauge transformation.}. There does not exist an analytical standardized procedure to calculate the $N$ derivatives: it could be an easy job for some models but, especially for multi-field inflationary scenarios, the calculation could be a cumbersome task \cite{Vernizzi:2006ve}.

\section{\label{secD}Correlators and the non-gaussianity parameter $f_{NL}$}
The inverse Fourier transform of the curvature perturbation is defined as:
\begin{equation}
\zeta(\vec{x},t) = \int \frac{d^3 k}{(2\pi)^3} e^{i\vec{k}\cdot\vec{x}} \zeta(\vec{k},t) .\label{secD_12}
\end{equation}
The inverse Fourier transform of each field perturbation is equally defined as:
\begin{equation}
\delta\varphi_i(\vec{x},t) = \int \frac{d^3 k}{(2\pi)^3} e^{i \vec{k}\cdot\vec{x}} \delta\varphi_i(\vec{k},t).\label{secD_13}
\end{equation}
These definitions allow us to express the two-point and three-point correlation functions of $\zeta$ in the following way:
\begin{eqnarray}
&& \left<\zeta(\vec{k_1},t)\zeta(\vec{k_2},t)\right> = N_i N_j \left<\delta\varphi_i(\vec{k_1},t)\delta\varphi_j(\vec{k_2},t)\right>, \nonumber \\
&& \label{secD_14} \\
&& \left<\zeta(\vec{k_1},t)\zeta(\vec{k_2},t)\zeta(\vec{k_3},t)\right> = \nonumber \\
&& N_i N_j N_k \left<\delta\varphi_i(\vec{k_1},t)\delta\varphi_j(\vec{k_2},t)\delta\varphi_k(\vec{k_3},t)\right>, \label{secD_14_1}
\end{eqnarray}
where the $N$ derivatives are calculated so that $t_1 = t_2$.
The latter expressions are valid only if the second-order perturbations are negligible. If the correlation functions of $\zeta$ are not gaussian, the three-point correlator of $\zeta$ is the first quantity that encodes the non-gaussian behaviour \cite{Linde} through the expression:
\begin{eqnarray}
\left\langle \zeta(\vec{k}_{1},t)\zeta(\vec{k}_{2},t)\zeta(\vec{k}_{3},t) \right\rangle&=&(2\pi)^{3}\delta ^{3}(\vec{k}_{1}+\vec{k}_{2}+\vec{k}_{3})\nonumber\\
&& \times \ B_{\zeta}(k_{1},k_{2},k_{3}), \label{bsdef}
\end{eqnarray}
where the $B_\zeta$ function is known as the bispectrum.  The bispectrum can in turn be expressed in terms of the power spectrum $P_\zeta$ of the primordial curvature perturbation in the following way:
\begin{eqnarray}
B_{\zeta}(k_{1},k_{2},k_{3}) &=& \frac{6}{5} f_{NL} \Big[P_\zeta(k_1) P_\zeta(k_2) + cyclic \nonumber \\
&& permutations \Big].
\end{eqnarray} 
The power spectrum $P_\zeta$ is, of course, defined from the two-point correlator of $\zeta$ in a similar way as the bispectrum is defined from the three-point correlator:
\begin{equation}
\left\langle \zeta(\vec{k}_{1},t)\zeta(\vec{k}_{2},t) \right\rangle = (2\pi)^{3}\delta ^{3}(\vec{k}_{1}+\vec{k}_{2}) \ P_{\zeta}(k). \label{psdef}
\end{equation}
When combined together, the expressions in Eqs. (\ref{bsdef}) - (\ref{psdef}) lead to
\begin{equation}
f_{NL} \approx \frac{5}{18}\frac{\langle \zeta\zeta\zeta \rangle }{\langle \zeta\zeta \rangle^2}, \label{secD_15}
\end{equation}
where the correlators in the latter expression are calculated in the real space.  Thus, we can easily recognize that the knowledge of the $N$ derivatives and the correlators of the field perturbations is fundamental for the calculation of the non-gaussianity parameter $f_{NL}$.

\section{\label{secE}Slow roll and the extended slow-roll conditions}

Taking the background definition of the energy density $\rho(t)$ from Eq. (\ref{secB_3}) and the background equation of motion for the scalar fields from Eq. (\ref{secB_6}), it is possible to describe the conditions during the inflationary mechanism, maintaining the Hubble parameter frozen until the universe has reached at least 62 e-folds of expansion, with the slow-roll conditions \cite{Linde,Liddle,Weinberg} for the background quantities:
\begin{eqnarray}
\dot\varphi_i(t)^2 &\ll& V(\varphi_i(t)),\label{secE_1} \\
|\ddot\varphi_i(t)| &\ll& |3H\dot\varphi_i(t)|.\label{secE_2}
\end{eqnarray}
When the perturbations are taken into account through the Eqs. (\ref{secB_9}) and (\ref{secB_12}), the implicit dependence on the position $\vec{x}$ makes the slow-roll expressions be ``extended'' for describing the inflationary period: 
\begin{eqnarray}
\dot\varphi_i(\vec{x},t)^2 &\ll& V(\varphi_i(\vec{x},t)), \label{secE_3} \\
|\ddot\varphi_i(\vec{x},t)| &\ll& |3 H(\vec{x},t) \dot\varphi_i(\vec{x},t)|. \label{secE_4}
\end{eqnarray}
We have to remember that a cosmological quantity is composed of two pieces: the homogeneous and isotropic piece that only depends on time, and the perturbation that depends on time and space:
\begin{eqnarray}
&& \left(\dot\varphi_i(t)+\dot\delta\varphi_i(\vec{x},t)\right)^2 = \nonumber \\
&& \dot\varphi_i(t)^2 + 2\dot\varphi_i(t)\delta\dot\varphi_i(\vec{x},t)+{\delta\dot\varphi_i(\vec{x},t)}^2. \label{secE_5}
\end{eqnarray}
Ignoring second-order contributions leads thereby to:
\begin{equation}
\dot\varphi_i(t)^2 + 2\dot\varphi_i(t) \delta\dot\varphi_i(\vec{x},t) \ll V\left(\varphi_i(t) + \delta\varphi_i(\vec{x},t)\right). \label{secE_6}
\end{equation}
The latter expression may reproduce the original (background) slow-roll condition of Eq. (\ref{secE_1}) only if the following set of conditions are satisfied:
\begin{eqnarray}
\dot\varphi_i(t) &\gg& \delta\dot\varphi_i(\vec{x},t), \label{secE_7} \\
\varphi_i(t) &\gg& \delta\varphi_i(\vec{x},t). \label{secE_8}
\end{eqnarray}
Regarding the second slow-roll condition in Eq. (\ref{secE_2}), it is expressed as:
\begin{eqnarray}
&& |\ddot\varphi_i(t) + \ddot\delta\varphi_i(\vec{x},t)| \ll \nonumber \\
&& 3 |\left(H(t)+\delta H(\vec{x},t)\right)\left(\dot\varphi_i(t)+\delta\dot\varphi_i(\vec{x},t)\right)|, \label{secE_9}
\end{eqnarray}
when introducing the perturbations. 
The latter expression may reproduce the original slow-roll condition described in Eq. (\ref{secE_2}) only if the following set of conditions are satisfied:
\begin{eqnarray}
\ddot\varphi_i(t) &\gg& \delta\ddot\varphi_i(\vec{x},t), \label{secE_10} \\
H(t) &\gg& \delta H(\vec{x},t). \label{secE_11}
\end{eqnarray}
The set of conditions depicted in Eqs. (\ref{secE_3}) and (\ref{secE_4}) are what we call the \emph{extended slow-roll conditions}.  They include the background slow-roll conditions of Eqs. (\ref{secE_1}) and (\ref{secE_2}) but go beyond them.

\section{\label{secF}Calculating the $N$ derivatives}
Employing the definitions of the global amount of expansion $N$ and Hubble parameter $H$, and making use also of the background slow-roll conditions, it is possible to rewrite $N$ as:
\begin{equation}
d N = - \frac{V}{3 m_p^2} \frac{d\varphi_i}{\frac{\partial V}{d\varphi_i }}. \label{secF_3}
\end{equation}
The \emph{local} version of the latter equation, \emph{which requires the application of the extended slow-roll conditions}, leads to the following expression:
\begin{equation}
N = - \frac{1}{m_p^2}\int \frac{V}{V'} d\varphi_i, \label{secF_4} 
\end{equation}
where the denominator of the integrand represents a field derivative and the limits of integration correspond to field values evaluated at the same time ($t_1 = t_2$) but on different slicings:  the lower limit is evaluated on the flat slice and the upper limit is evaluated in the uniform energy density slice. With this expression for the local amount of expansion $N$, and taking into account the procedure followed in Ref. \cite{Mulryne:2009kh}, the $N$ derivatives for a two-field inflation model are obtained:
\begin{eqnarray}
N_1 &=& V_1 V, \label{secF_5} \\
N_2 &=& V_2 V, \label{secF_6} \\
N_{12} &=& 2\left(\frac{V}{V_1}\right)\left(\frac{V}{V_2}\right)(V_1)^2 \nonumber\\
&& \times \left[\frac{-V_{11}}{V}(V_2)^2 + \frac{V_{12} V}{V_2 V}(V_2)^4 \right] \nonumber\\ 
&& +2\left(\frac{V}{V_1}\right)\left(\frac{V}{V_2}\right)(V_1)^2\nonumber\\
&& \times \left[\left(\frac{V_{11}}{V}-\frac{V_{22}}{V}+\frac{V_2 V_{12}}{V_1 V} \right)(V_1)^2 (V_2)^2 \right], \nonumber \\
&& \label{secF_7} \\
N_{11} &=& \left(1-\frac{V_{11} V}{V_1^2} \right)(V_1)^2 + 2\left(\frac{V}{V_1}\right)^2(V_2)^2 \nonumber\\ 
&& \times \left[\frac{V_{11}}{V}(V_2)^2+\frac{V_1 V_{12}}{V V_2}(V_2)^4\right] \nonumber\\
&& + 2\left(\frac{V}{V_1}\right)^2(V_2)^2 \nonumber\\
&& \times \left[\left(\frac{V_{11}}{V} -\frac{V_{22}}{V} + \frac{V_2 V_{12}}{V V_1} \right)(V_1)^2 (V_2)^2 \right], \nonumber \\
&& \label{secF_8} \\
N_{22} &=& \left(1-\frac{V_{22} V}{V_2^2} \right)(V_2)^2 + 2\left(\frac{V}{V_2}\right)^2(V_1)^2 \nonumber\\ 
&& \times \left[\frac{V_{22}}{V}(V_1)^2+\frac{V_2 V_{21}}{V V_1}(V_2)^4\right] \nonumber\\
&& + 2\left(\frac{V}{V_2}\right)^2(V_1)^2 \nonumber\\
&& \times \left[\left(\frac{V_{22}}{V} -\frac{V_{11}}{V} + \frac{V_1 V_{21}}{V V_2} \right)(V_2)^2 (V_1)^2 \right], \nonumber \\
&& \label{secF_8_1}
\end{eqnarray}
where the subindices mean derivatives with respect to the first or second scalar field.
Now, truncating the $\delta N$ series expansion of Eq. (\ref{secC_8}) at second order, we can calculate the two-point and three-point correlation functions of $\zeta$:
\begin{eqnarray}
\langle \zeta\zeta \rangle &=& N_1^2 \Sigma_{11} + N_2^2 \Sigma_{22} + 2 N_1 N_2 \Sigma_{12},\label{secF_9} \\
\langle \zeta\zeta\zeta \rangle &=& N_1^3 \alpha_{111} + N_2^3 \alpha_{222} + 3 N_1^2 N_2 \alpha_{112} \nonumber \\
&& + 3 N_1 N_2^2 \alpha_{122}+\frac{9}{2} N_{1}^2 N_{11} \Sigma_{11}^2 \nonumber \\
&& + \frac{9}{2} N_{2}^2 N_{22} \Sigma_{22}^2 \nonumber \\
&& + 9 \left( N_{1} N_{2} N_{11} + N_{1}^2 N_{12} \right) \Sigma_{11} \Sigma_{12} \nonumber \\
&& + \frac{3}{2} \left( N_{2}^2 N_{11} + N_{1}^2 N_{22} + 4 N_{1} N_{2} N_{12} \right) \nonumber \\
&& \times \left(\Sigma_{11} \Sigma_{22} + 2 \Sigma_{12}^2 \right) \nonumber \\
&& - \frac{3}{2} \left( N_{11} \Sigma_{11} + 2 N_{12} \Sigma_{12} + N_{22} \Sigma_{22}
\right) \langle \zeta \zeta \rangle, \nonumber \\
&& \label{secF_10}
\end{eqnarray}
where the quantities $\Sigma_{ij}$ and $\alpha_{ijk}$ represent the correlation functions of the field perturbations:
\begin{eqnarray}
\Sigma_{ij} = \langle \delta\varphi_i(\vec{x},t)\delta\varphi_j(\vec{x},t)\rangle,\\\label{secF_11}
\alpha_{ijk} = \langle \delta\varphi_i(\vec{x},t)\delta\varphi_j(\vec{x},t)\delta\varphi_k(\vec{x},t)\rangle, \label{secF_12}
\end{eqnarray}
evaluated at the time when $\zeta$ is to be calculated.  Since we know these correlators at the time of horizon exit \cite{Mukhanov,Seery:2005gb}, we need to evolve them to the desired time when $\zeta$ is to be calculated.
The next section will explain the way to obtain the time evolution of the correlators of the field perturbations.

\section{\label{secG}Moment transport equations}
As was proposed, the main idea of this methodology is to avoid the time evolution of the $N$ derivatives, transporting that evolution to the correlation functions of the field perturbations. 

The statistical properties of the studied quantities are encoded in the probability density function (PDF) $P(\varphi_i)$. That PDF only needs to exhibit useful information for the first three moments: the mean, the variance and the skewness. This fact allows us to obtain the $f_{NL}$ parameter. Higher odd moments will be assumed to vanish and higher even moments will be obtained as products of the variance. The only difference between the gaussian distribution and the nearly-gaussian distribution we will employ is the third-order moment, the skewness.

One suitable PDF for that purpose is, for the one-field case:
\begin{equation}
\label{e:P1D}
P(\varphi)=P_g(\varphi) \times P_{ng}(\varphi) ,\label{secG_1}
\end{equation}
where $P_g$ is a pure gaussian distribution function. $P_{ng}$ represents the non-gaussian part of the distribution. In the one-field case, following a proper cumulant expansion process, $P_{ng}$ takes the form \cite{Mulryne:2009kh,Blinnikov:1997jq}:
\begin{equation}
P_{ng}(\varphi)=\left[1 + \frac{\alpha}{6 \sigma^3} H_3\left(\frac{\varphi-\varphi_0}{\sigma}\right)\right], \label{secG_3}
\end{equation}
where the $H_3$ term is a Chevishev-Hermite polynomial of grade three, $\varphi$ is the field, $\varphi_0$ is the mean and $\sigma$ is the standard deviation. 

The probability is a conserved quantity and then it has associated a continuity equation:
\begin{equation}
\frac{\partial P(\varphi)}{\partial N} + \frac{\partial}{\partial\varphi}\left({u(\varphi)P(\varphi)}\right)=0, \label{secG_8} 
\end{equation}
where the time parameter is replaced by $N$ which is a physically more significant quantity in the primordial inflationary epoch.  The velocity field is represented by $u(\varphi)$.
The Eq. (\ref{secG_8}) can be obtained directly from the study of the kinematics of heavy molecules immersed in an almost non-interacting bath of light molecules \cite{Landau}, neglecting the second-order derivatives that represent the diffusion term. Another way to obtain the probability conservation equation is through the stochastic processes approach that leads to the Fokker-Planck equation \cite{Jacobs}.

From the definition of the slow-roll parameter $\epsilon$ \cite{Linde,Liddle,Weinberg}:
\begin{equation}
	\epsilon=-\frac{\dot{H}}{H^2}, \label{secG_5}
\end{equation}
and the equation of motion for the field, Eq. (\ref{secB_12}), we can write:
\begin{equation}
	u(\varphi)=\frac{1}{m_p^2}\frac{\partial(ln V)}{\partial\varphi}. \label{secG_6}
\end{equation}
Expanding the velocity field centered in the mean value of $\varphi$, i.e. $\varphi_0$, it yields:
	\begin{equation}
		u(\varphi) = u_0	+ u_{\varphi} (\varphi - \varphi_0) + \frac{1}{2} u_{\varphi\varphi} (\varphi - \varphi_0)^2+ \cdots ,
	\end{equation}
where the subindices of the $u-$terms mean zeroth-, first-, and second-order derivatives with respect to the $\varphi$ field. 

From the continuity equation that describes the probability conservation, Eq. (\ref{secG_8}), it is possible to replace the definition of the nearly-gaussian PDF from Eqs. (\ref{secG_1}) and (\ref{secG_3}) and obtain the set of equations:
		\begin{eqnarray}
		\frac{d \varphi_0}{d N} &=&	u_0 + \frac{1}{2} u_{\varphi\varphi} \sigma^2, \\
		\label{e:EvoVar1D}
		\frac{d \sigma^2}{d N} &=& 2 u_{\varphi} \sigma^2
		+ u_{\varphi\varphi} \alpha ,\\
		\label{e:EvoAlpha1D}
		\frac{d \alpha}{d N} &=& 3 u_{\varphi} \alpha + 3 u_{\varphi\varphi} \sigma^4.
	\end{eqnarray}
Extending the definition of the Eq. (\ref{secG_8}) to the two-field case:
\begin{equation}
\frac{\partial P(\varphi_i)}{\partial N} + \frac{\partial}{\partial\varphi_i}\left({u(\varphi_j)P(\varphi_j)}\right)=0, \label{secG_7} 
\end{equation}
and using the extended version of the two-field probability density distribution, the following set of moment transport equations are obtained:
		\begin{eqnarray}
			\label{etm1}
			\frac{d \Phi_i}{d N} &=& u_{i0}+ \frac{1}{2} \Sigma_{jk} u_{ijk} , \\
			\frac{d \Sigma_{ij}}{d N} &=& u_{ik} \Sigma_{kj}
			+ u_{jk} \Sigma_{ki} \nonumber \\
			&& + \frac{1}{2} \left(\alpha_{imn} u_{jmn} +	\alpha_{jmn} u_{imn} \right) , \label{etm2}  \\
			\frac{d \alpha_{ijk}}{d N}
			&=&
			u_{in} \alpha_{njk} + \Sigma_{jm} u_{imn} \Sigma_{nk} \nonumber \\
			&&+ \textrm{cyclic permutations $i \rightarrow j \rightarrow k$},
			\label{etm3}
		\end{eqnarray}
where $\Phi_i$ means the central value of the $\varphi_i$ field.
As can be seen, the set of equations shows a time evolution of the first three moments of the distribution function. The original idea of carrying the problem of the time evolution from the $N$ derivatives to the correlation functions of the field perturbation is accomplished. 

\section{\label{secH}Numerical Implementation}
Solving the set of ordinary differential equations (ODE's) presented in Eqs. (\ref{etm1}) - (\ref{etm3}) will be necessary in order to obtain $f_{NL}$, following the definition presented in Eq. (\ref{secD_15}).
The set of moment transport equations, derived from the probability conservation, may be solved numerically for each inflationary model. 

The numerical implementation was implemented for the double quadratic inflation, previously analyzed in the Ref. \cite{Vernizzi:2006ve} where the  potential is defined by:
\begin{equation}
V(\varphi,\chi)= \frac{1}{2}m_\varphi^2 \varphi^2+\frac{1}{2}m_\chi^2 \chi^2. \label{vmodel}
\end{equation}
Employing the initial conditions provided in the Ref. \cite{Dias:2011xy}:
\begin{itemize}
\item  $m_\varphi = 9 \times 10^{-5} \ m_p$,
\item  $m_\chi = 1 \times 10^{-5} \ m_p$,
\item  $\varphi_0 = 8.9 \ m_p$,
\item  $\chi_0 = 12.9 \ m_p$,
\end{itemize}
and implementing the numerical code for solving the system of differential equation composed by:
\begin{itemize}
\item  two ODE's for the first moment of the PDF, as shown in the Eq. (\ref{etm1}),
\item  three ODE's for the second moment of the PDF, as shown in the Eq. (\ref{etm2}),
\item  four ODE's for the third moment of the PDF, as shown in the Eq. (\ref{etm3}),
\end{itemize}
the evolution of each correlation function is obtained. Replacing these functions in the Eqs. (\ref{secF_9}) and (\ref{secF_10}), and calculating the $N$ derivatives from Eqs. (\ref{secF_5}) - (\ref{secF_8_1}) and (\ref{vmodel}), the two-point and three-point correlation functions of the primordial curvature perturbation $\zeta$ are obtained.
\begin{figure}
\includegraphics[scale=0.5]{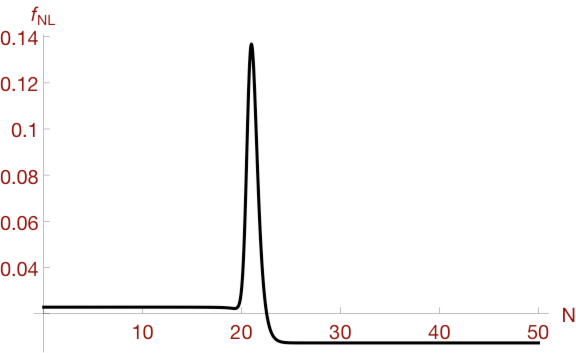}
\caption{Primordial non-gaussianity evolution in the two-field model of Eq. (\ref{vmodel}). Graphic extracted from Ref. \cite{Dias:2011xy}.} \label{notfound}
\end{figure}
The resulting graphic in Fig. \ref{notfound} shows the evolution of the $f_{NL}$ parameter for the selected inflationary mechanism; the nongaussianity in this model exhibits a spike around $N = 20$ but it is anyway small enough to be observable \cite{Ade:2015ava}.

\section{Conclusions}
The idea of carrying the problem of the time evolution from the amount of expansion derivatives to the correlation functions of the field perturbations is the fundamental basis of the moment transport methodology. Why is it necessary? Because there does not exist a standardized procedure of calculating analytically the $N$ derivatives for general inflationary models.

The studied methodology has a well defined procedure:
\begin{enumerate}
	\item Select your scalar inflationary model.
	\item Impose the extended slow-roll conditions, expressed in Eqs. (\ref{secE_3}) - (\ref{secE_4}).
	\item Employ the Eq. (\ref{secB_12}) to obtain the evolution of the scalar fields.
	\item Use the moment transport equations, Eqs. (\ref{etm1}) - (\ref{etm3}), to obtain the evolution of the correlation functions of the field perturbations starting from their values at horizon exit reported in Refs. \cite{Mukhanov,Seery:2005gb}.
	\item Calculate the $N$ derivatives using the Eqs. (\ref{secF_5}) - (\ref{secF_8_1}).
	\item Obtain the correlators $\langle\zeta\zeta\rangle$ and $\langle\zeta\zeta\zeta\rangle$ by means of the Eqs. (\ref{secF_9}) - (\ref{secF_10}).
	\item Calculate the non-gaussianity $f_{NL}$ parameter employing Eq. (\ref{secD_15}).
\end{enumerate}

This algorithm can be applied \emph{only if the extended slow roll conditions are satisfied} by the cosmological model of interest. Otherwise, making use of the expression in Eq. (\ref{secF_4}), and therefore of Eqs. (\ref{secF_5}) - (\ref{secF_8_1}), would lead to results which go beyond the regime of applicability of the methodology employed (see for example the differences between Refs. \cite{Byrnes:2008zy} and \cite{Cogollo:2008bi}). 


For the cosmological inflationary model studied in Ref. \cite{Mulryne:2009kh}, the evolution of the $f_{NL}$ parameter (see Fig. \ref{notfound}) is in agreement with the $\delta N$ formalism as can be observed in Refs. \cite{Mulryne:2009kh,Vernizzi:2006ve}.  The resulting nongaussianity is very small, in agreement with theoretical expectations \cite{Linde} and with the latest satellite results \cite{Ade:2015ava}.

\vspace{5mm}
{\it Acknowledgements.}
This work was supported by COLCIENCIAS grant number 110656933958 RC 0384-2013, COLCIENCIAS - ECOS-NORD grant number RC 0899-2012 with the help of ICETEX, and by Fundaci\'on para la Promoci\'on de la Investigaci\'on y la Tecnolog\'ia del Banco de la Rep\'ublica (COLOMBIA) grant number 3025 CT-2012-02.  C. A. S. acknowledges support for mobility from VIE (UIS).  This work is entirely based on Ref. [4] and is the result of C. A. S.'s MSc thesis.


\begin{thebibliography}{99}

\bibitem{Linde}
	D. H. Lyth and A. R. Liddle, The primordial density perturbation: cosmology, inflation and the origin of structure, Cambridge University Press, 
	(2009).
	
\bibitem{Liddle}
	A. R. Liddle, An introduction to modern cosmology, John Wiley \& Sons, (2003).	
	
\bibitem{Weinberg}
	S.~Weinberg,
 Cosmology, Oxford University Press, (2008).

\bibitem{Mulryne:2009kh} 
	D. J. Mulryne, D.Seery and D.Wesley,
  JCAP {\bf 1001}, 024 (2010).

\bibitem{Mulryne:2010rp} 
  D.~J.~Mulryne, D.~Seery and D.~Wesley,
  JCAP {\bf 1104}, 030 (2011).

	\bibitem{Mulryne:2013oba} 
  D.~J.~Mulryne,
  arXiv:1302.4636 [astro-ph.CO].

\bibitem{Sasaki:1995aw}
	M.~Sasaki and E.~D. Stewart, 
	Prog. Theor. Phys. {\bf 95}, 71 (1996).

\bibitem{Lyth:2004gb}
  D.~H.~Lyth, K.~A.~Malik and M.~Sasaki,
  JCAP {\bf 0505}, 004 (2005).

\bibitem{Lyth:2005fi}
  D.~H.~Lyth and Y.~Rodr\'iguez,
  Phys.\ Rev.\ Lett.\  {\bf 95}, 121302 (2005).

\bibitem{Hinshaw:2012aka} 
	G.~Hinshaw {\it et al.}  [WMAP Collaboration],
  Astrophys.\ J.\ Suppl.\ Ser.\  {\bf 208}, 19 (2013).

\bibitem{Ade:2013ydc} 
  P.~A.~R.~Ade {\it et al.}  [Planck Collaboration],
  arXiv:1502.01593 [astro-ph.CO].

  
\bibitem{Vernizzi:2006ve}
	F.~Vernizzi and D.~Wands, 
  JCAP {\bf 0605}, 019 (2006).
 
 
\bibitem{Mukhanov}
V. F. Mukhanov and S. Winitzki, Introduction to quantum effects in gravity, Cambridge University Press, (2007).
 
\bibitem{Seery:2005gb} 
  D.~Seery and J.~E.~Lidsey,
  JCAP {\bf 0509}, 011 (2005).
 


\bibitem{Blinnikov:1997jq}
	S.~Blinnikov and R.~Moessner,
	Astron. Astrophys. Suppl. Ser. {\bf 130}, 193 (1998).

\bibitem{Landau}
E. M. Lifshitz and L. P. Pitaevskii­, Physical kinetics, Course of theoretical physics,  Butterworth-Heinemann, (1999).

\bibitem{Jacobs}
K. Jacobs, Stochastic processes for physicists:  understanding noisy systems,  
 Cambridge University Press, (2010).

\bibitem{Dias:2011xy} 
  M.~Dias and D.~Seery,
  Phys.\ Rev.\ D {\bf 85}, 043519 (2012).
  
\bibitem{Ade:2015ava} 
  P.~A.~R.~Ade {\it et al.}  [Planck Collaboration],
  arXiv:1502.01592 [astro-ph.CO].

\bibitem{Byrnes:2008zy} 
  C.~T.~Byrnes, K.~Y.~Choi and L.~M.~H.~Hall,
  JCAP {\bf 0902}, 017 (2009).
  
\bibitem{Cogollo:2008bi} 
  H.~R.~S.~Cogollo, Y.~Rodr\'iguez and C.~A.~Valenzuela-Toledo,
  JCAP {\bf 0808}, 029 (2008).

\end{thebibliography}
\end{document}